\documentclass[smallextended]{svjour3}
\usepackage[T1]{fontenc}
\usepackage[latin1]{inputenc}
\usepackage{graphicx}
\usepackage{amssymb}
\usepackage{amsmath}
\usepackage{wasysym}
\usepackage{natbib}
\usepackage{lineno}
\usepackage[usenames]{color}
\usepackage{pdflscape}
\usepackage{rotating}
\usepackage{soul}
\usepackage{xcolor, colortbl}
\definecolor{Gray}{gray}{0.9}
\usepackage{etoolbox}
\usepackage[letterpaper,margin=2.5cm]{geometry}

\usepackage[pdftex,colorlinks,linkcolor=blue,anchorcolor=blue,urlcolor=blue,citecolor=blue]{hyperref}

\newtoggle{supplement}
\toggletrue{supplement}

\newtoggle{removemanunobox}
\toggletrue{removemanunobox}

\iftoggle{removemanunobox}
{
\makeatletter
\renewcommand\makeheadbox{{%
\hbox to0pt{\vbox{\baselineskip=10dd\hbox
to\hsize{\kern3pt\vbox{\kern12pt
\hbox{\ }
\hbox{Pre-print}
\hbox{Published as: \emph{Climatic Change}, doi:10.1007/s10584-015-1451-x}
}}}%
}}}
\makeatother
}

\smartqed
\newcommand{\about}{\mathord{\sim}}

\journalname{Climatic Change}

\begin{document}

\title{Past and future sea-level rise along the coast of North Carolina, USA }
\author{Robert E. Kopp, Benjamin P. Horton, Andrew C. Kemp and Claudia Tebaldi}

\institute{
R. E. Kopp \at Department of Earth \& Planetary Sciences, Rutgers Energy Institute, and Institute of Earth, Ocean, \& Atmospheric Sciences, Rutgers University, New Brunswick, NJ, USA. \email{robert.kopp@rutgers.edu} \and B. P. Horton \at Sea Level Research, Department of Marine \& Coastal Sciences and Institute of Earth, Ocean, \& Atmospheric Sciences, Rutgers University, New Brunswick, NJ, USA and Earth Observatory of Singapore and Asian School of the Environment, Nanyang Technological University, Singapore \and  A. C. Kemp \at Department of Earth \& Ocean Sciences, Tufts University, Medford, MA, USA \and C. Tebaldi \at National Center for Atmospheric Research, Boulder, CO, USA}

\date{Received: 31 October 2014. Accepted: 8 June 2015.\\
The final publication is available at \url{http://dx.doi.org/10.1007/s10584-015-1451-x}}

\maketitle
\vspace{-12pt}

\begin{abstract}
We evaluate relative sea level (RSL) trajectories for North Carolina, USA, in the context of tide-gauge  measurements and geological sea-level  reconstructions spanning the last $\about$11,000 years. RSL rise was fastest ($\about$7 mm/yr) during the early Holocene and slowed over time with the end of the deglaciation. During the pre-Industrial Common Era (i.e., 0--1800 CE), RSL rise ($\about$0.7 to 1.1 mm/yr) was driven primarily by glacio-isostatic adjustment, though dampened by tectonic uplift along the Cape Fear Arch. Ocean/atmosphere dynamics caused centennial variability of up to $\about$0.6 mm/yr around the long-term rate. It is extremely likely (probability $P = 0.95$) that 20th century RSL rise at Sand Point, NC, (2.8 $\pm$ 0.5 mm/yr) was faster than during any other century in at least 2,900 years. Projections based on a fusion of process models, statistical models,  expert elicitation, and expert assessment indicate that RSL at Wilmington, NC, is very likely ($P = 0.90$) to rise by 42--132 cm between 2000 and 2100 under the high-emissions RCP 8.5 pathway.  Under all emission pathways, 21st century RSL rise is very likely ($P > 0.90$) to be faster than during the 20th century.    Due to RSL rise, under RCP 8.5,  the current `1-in-100 year' flood  is expected  at Wilmington  in $\about$30 of the 50 years between 2050-2100.
\end{abstract}

\section{Introduction}

Sea-level rise threatens coastal populations, economic activity, static infrastructure, and ecosystems by increasing the frequency and magnitude of flooding in low-lying areas.  For example, Wilmington, North Carolina (NC), USA,  experienced nuisance flooding $\about$2.5 days/yr on average between 1938 and 1970, compared to 28 days/yr between 1991 and 2013 \citep{Ezer2014a}. However, the likely magnitude of 21st century sea-level rise -- both globally and regionally -- is uncertain. Global mean sea-level  (GMSL) trends are driven primarily by ocean heat uptake and land ice mass loss. Other processes, such as ocean dynamics, the static-equilibrium `fingerprint' effects of land ice loss on the height of Earth's geoid and  surface, tectonics, and glacio-isostatic adjustment (GIA), are spatially variable and cause sea-level rise to vary in rate and magnitude between regions \citep{Milne2009a,Stammer2013a}. Sound risk management necessitates that decision-makers tasked with creating resilient coastal ecosystems, communities, and economies  are informed by reliable projections of the risks of regional  relative sea-level (RSL) change (not just GMSL change) on policy-relevant (decadal) timescales \citep{Poulter2009a}. 

The North Carolina Coastal Resources Commission (CRC)'s Science Panel on Coastal Hazards (\citeyear{NCCRC2010a})  recommended the use of 1 m of  projected sea-level rise between 2000 and 2100 for statewide policy and planning purposes in North Carolina. Since the CRC's 2010  assessment, several advances have been made in the study of global and regional sea-level change. These include new reconstructions of sea level in the U.S. generally and North Carolina in particular during the Holocene (the last $\about11.7$ thousand years) \citep{Engelhart2012a,vandePlassche2014a} and the Common Era (the last two millennia) \citep{Kemp2011a,Kemp2013a,Kemp2014a},  estimates of 20th century GMSL change \citep{Church2011a,Ray2011a,Hay2015a},  localized projections of future sea-level change \citep{Kopp2014a}, and  state-level assessments of the cost of sea-level rise \citep{Houser2015a}. 

Political opposition led to North Carolina House Bill 819/Session Law 2012-202, which blocked the use of the 1 m projection for regulatory purposes and charged the Science Panel on Coastal Hazards to deliver an updated assessment in 2015 that considered \emph{``the full range of global, regional, and North Carolina-specific sea-level change data and hypotheses, including sea-level fall, no movement in sea level, deceleration of sea-level rise, and acceleration of sea-level rise''} \citep{NCHB819_2011}. Here, we assess the likelihood of these  trajectories with respect to past and future sea-level changes in North Carolina.

\begin{figure}[btp]
   \centering
  \includegraphics[width=12cm]{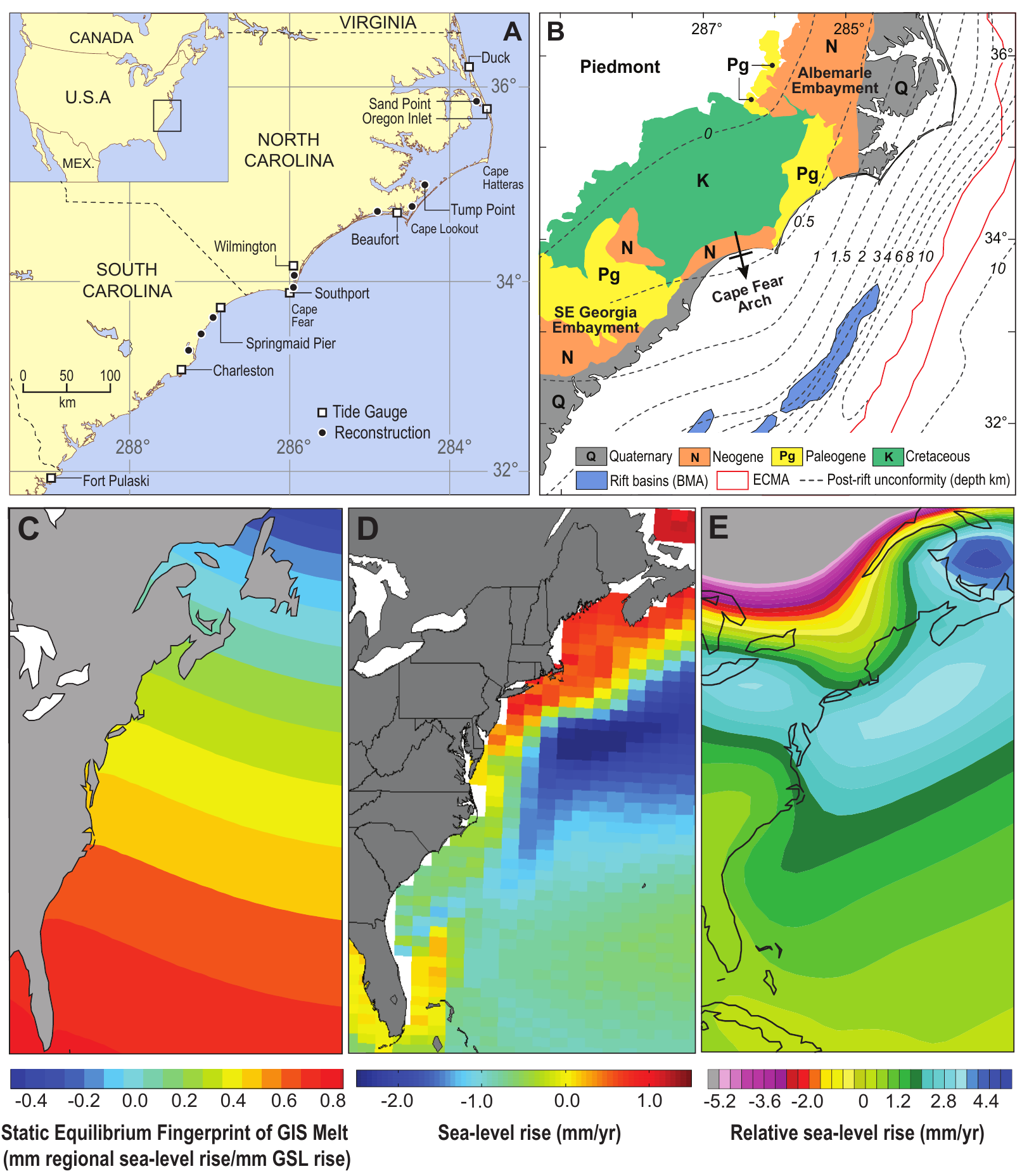}

   \caption{(A) Location map. (B) Map of regional shallow subsurface geology, post-rift unconformity, and large-scale structural geology \citep{Dillon1988a,Gohn1988a,Grow1988a,NCGS2004a}. (C)  Static-equilibrium fingerprint of RSL change from uniform melting of the Greenland Ice Sheet \citep{Mitrovica2011a}, in units of mm RSL rise per mm GMSL rise. (D) Ocean dynamic contribution to RSL over 2006-2100 in the Community Earth System Model RCP 8.5 experiment from the Coupled Model Intercomparison Project Phase 5 \citep{Taylor2012a}. (E) GIA contribution to RSL under the ICE-6G VM5b model \citep{Engelhart2011a}} \label{fig:map} 
\end{figure}

\section{Mechanisms for global, regional, and local relative sea-level changes}

Relative sea level (RSL) is  the difference in elevation between the solid Earth surface and the sea surface at a specific location and point in time.  Commonly, it is time-averaged to minimize the influence of tides and is compared to the present as the reference period \citep{Shennan2012a}. RSL averaged over all ocean basins yields an estimate of GMSL.

GMSL rise is driven primarily by (1) increases in ocean mass due to melting of land-based  glaciers \citep[e.g.,][]{Marzeion2012a} and ice sheets \citep[e.g.,][]{Shepherd2012a} and  (2) expansion of ocean water as it warms \citep[e.g.,][]{Gregory2010a}. Changes in land water storage due to dam construction and groundwater withdrawal also contributed to 20th century GMSL change \citep[e.g.,][]{Konikow2011a}. RSL differs from GMSL because of (1) factors causing vertical land motion, such as tectonics, sediment compaction, and groundwater withdrawal; (2) factors affecting both the height of the solid Earth and the height of Earth's geoid, such as long-term GIA and the more immediate `sea-level fingerprint' static-equilibrium response of the geoid and the solid Earth to redistribution of mass between land-based ice and the ocean; and (3) oceanographic and atmospheric factors affecting sea-surface height relative to the geoid, such as changes in ocean-atmospheric dynamics and the distribution of heat and salinity within the ocean \citep[e.g.,][]{Kopp2014a,Kopp2015a}

Along the U.S. Atlantic coast, the principal mechanism for regional departures from GMSL during the Holocene is GIA, which is the ongoing, multi-millennial response of  Earth's shape and geoid to large-scale changes in surface mass load \citep[e.g.,][]{Clark1978a} (Figure \ref{fig:map}e). Growth and thickening of the Laurentide ice sheet during the last glaciation  caused subsidence of land beneath the ice mass \citep{Clark2009a}. A compensating outward flow in the mantle created a peripheral bulge around the ice margin in the U.S. mid-Atlantic region. In addition to uplifting the solid Earth in the U.S. mid-Atlantic region, these flows also increased the regional height of the geoid and reduced the global volume of the ocean basin. These latter two factors led to a rising sea-surface height in the U.S. mid-Atlantic region and thus a total RSL fall less than the regional uplift \citep{Farrell1976a}. As the Laurentide ice sheet shrunk, mantle flow back toward the center of the diminishing ice sheet caused subsidence and progressive inward migration of the peripheral forebulge. One commonly used physical model of GIA (ICE-5G-VM2-90) yields contributions to 20th century sea-level rise of $\about$1.3 mm/yr at New York City and $\about$0.5 mm/yr at Wilmington, NC \citep{Peltier2004a}, but exact values depend upon assumptions regarding ice-sheet history and mantle viscosity. 

Along much of the U.S. Atlantic coast, the tectonic contribution to  RSL change is assumed to be negligible over timescales of centuries to millennia \citep[e.g.,][]{Rowley2013a}, but parts of the North Carolina coastal plain are underlain by the Cape Fear Arch  \citep{Sheridan1976a} (Figure \ref{fig:map}b). Geologic and geomorphic data suggest that uplift of the crest of the Cape Fear Arch began during the Pliocene \citep{Wheeler2006a} and is ongoing \citep{Brown1978a}. Late Holocene rates of uplift (RSL fall) have been estimated at $\about$0.2 $\pm$ 0.2 mm/yr \citep[e.g.,][]{Marple2004a,vandePlassche2014a}.

The static-equilibrium `fingerprint' contribution to RSL changes arises from the immediate response of Earth's geoid, rotation, and elastic lithosphere  to redistribution of mass between land ice and the ocean \citep{Clark1977a,Mitrovica2011a}. As the mass of an ice sheet or glacier shrinks, sea-level rise is greater in areas geographically distal to the land ice than in areas close to it, primarily because the gravitational attraction between the ice mass and the ocean is reduced. Greenland Ice Sheet (GrIS) mass loss, for instance, generates a meridional sea-level gradient along the U.S. Atlantic coast (Figure \ref{fig:map}c), where Maine experiences $\about$30\% of the global mean response, compared to $\about60\%$ in North Carolina and $\about80\%$ in south Florida. Melting of the West Antarctic Ice Sheet (WAIS), by contrast, causes a nearly uniform rise along the U.S. Atlantic coast (including North Carolina), which is about 20\% higher than the global average due primarily to the effect of WAIS mass loss on Earth's rotation \citep{Mitrovica2009a}. Though the magnitude of sea-level fingerprints proximal to a changing ice mass is sensitive to the internal distribution of that mass, this sensitivity diminishes  with distance. For example, at the distance of North Carolina, assumptions about the distribution of mass lost from GrIS have only an $\about 10\%$ effect on the fingerprint (i.e., a RSL effect equal to $\about 6\%$ of the global mean) \citep{Mitrovica2011a}.

Oceanographic effects change sea-surface height relative to the geoid \citep[e.g.,][]{Kopp2010a}. They include both global mean thermal expansion and regional changes in ocean-atmospheric dynamics and in the distribution of heat and salinity within the ocean. For example, changes in the Gulf Stream affect sea level in the western North Atlantic Ocean \citep[e.g.,][]{Kienert2012a,Ezer2013a}. As observed by satellite altimetry, the dynamic sea-surface height off of New Jersey averages $\about$60 cm lower than the height off of Bermuda. By contrast, off the North Carolina coast, the dynamic sea-surface height averages $\about30$ cm lower than off Bermuda, and this difference diminishes much more quickly off shore than it does north of  Cape Hatteras, where the Gulf Stream separates from the U.S. Atlantic coast and turns toward northern Europe \citep{Yin2013a}. Ocean modeling shows that a slower Gulf Stream, which can be caused by a weaker Atlantic Meridional Overturning Circulation or by shifting winds, would reduce these sea-level gradients, increasing sea level along the U.S. Atlantic coast north of Cape Hatteras  (Figure \ref{fig:map}d). A northward shift in the position of the Gulf Stream, which could result from a migration of the Intertropical Convergence Zone (ITCZ),  would similarly raise mid-Atlantic sea levels. In contrast, sea-surface height in coastal regions south of Cape Hatteras is less influenced by changes in the Gulf Stream \citep{Yin2013a}. 

Locally in North Carolina, RSL  also changes in response to  sediment compaction \citep{Brain2015a}, groundwater withdrawal \citep{Lautier2006a}, and tidal-range shifts.  North Carolina is partly located within the Albemarle Embayment (Figure \ref{fig:map}b), a Cenozoic depositional basin \citep{Foyle1997a} stretching from the Norfolk Arch at the North Carolina/Virginia border to southern Pamlico Sound at the Cape Lookout High. The embayment is composed of $\about 1.5$ km thick post-rift sedimentary rocks and Quaternary unconsolidated sediments \citep[e.g.,][]{Gohn1988a}, currently undergoing compaction \citep[e.g.,][]{vandePlassche2014a}.  

The influence of local factors on regional RSL reconstructions  is minimized by using proxy and instrumental data from multiple sites.  For example, \citet{Kemp2011a} concluded that local factors were not the primary driving mechanisms for  RSL change in  North Carolina over the last millennium, because the trends reconstructed at two sites located $\mathord{>}100$ km apart in  different water bodies closely agree.

\section{Methods}

\subsection{Historical reconstruction}

Tide gauges provide historic measurements of RSL for specific locations (Figure \ref{fig:map}a). In North Carolina, there are  two long-term tide-gauge records: Southport (covering 1933-1954, 1976-1988, and 2006-2007) and Wilmington (covering 1935 to present). Both have limitations: Southport has temporal gaps in the record, while the Wilmington record was influenced by deepening of the navigational channels, which increased the tidal range \citep{Zervas2004a}. There are also shorter records from Duck (1978 to present), Oregon Inlet (1977 and 1994 to present), and Beaufort (1953-1961, 1966-1967, and 1973 to present), which we also include in our analysis. 

Geological reconstructions provide proxy records of pre-20th century RSL. Our database of Holocene RSL reconstructions from North Carolina includes 107 discrete sea-level constraints from individual core samples collected at a suite of sites  \citep{Horton2009a,Engelhart2012a,vandePlassche2014a}. It also includes two continuous Common Era RSL reconstructions, from Tump Point (spanning the last $\about$1000 years) and Sand Point (spanning the last $\about$2000 years), produced using ordered samples from cores of salt-marsh sediment \citep{Kemp2011a} (Figure \ref{fig:map}a). Salt marshes from the U.S. Atlantic Coast provide higher-resolution reconstructions than other sea-level proxies (in North Carolina, $<0.1$ m vertically and $\pm$ 1 to $\pm$ 71 y geochronologically).  The combination of an extensive set of Holocene sea-level index points, multiple, high-resolution Common Era reconstructions, and tide-gauge measurements  makes North Carolina well suited to evaluating past sea-level changes.

We fit the proxy and tide-gauge observations to a spatio-temporal Gaussian process (GP) statistical model of the Holocene RSL history of the U.S. Atlantic Coast. The model is similar to that of \citet{Kopp2013a}, though with a longer temporal range and with geochronological uncertainty accommodated through the noisy-input GP method of \citet{McHutchon2011a}. To provide regional context, the fitted data also include records from outside of North Carolina, in particular salt-marsh reconstructions from New Jersey \citep{Kemp2013a} and Florida \citep{Kemp2014a} and all U.S. Atlantic Coast tide-gauge records in the \citet{PSMSL2014a} database with $\mathord{>}60$ years of data. To aid comparison with the proxy reconstructions, tide-gauge measurements were incorporated into the analysis as  decadal averages. The GP model represents sea level as the sum of spatially-correlated low-frequency (millennial), medium-frequency (centennial) and high-frequency (decadal) processes. Details are provided in the Supporting Information. All estimated rates of past RSL change  in this paper are based on application of the GP model to the combined data set and are quoted with $2 \sigma$ uncertainties. 

\subsection{Future projections}

Several data sources are available to inform sea-level projections, including process models of ocean and land ice behavior \citep[e.g.,][]{Taylor2012a,Marzeion2012a}, statistical models of local sea-level processes \citep{Kopp2014a}, expert elicitation on ice-sheet responses \citep{Bamber2013a} and expert assessment of the overall sea-level response \citep{IPCCAR5WG1ch13,Horton2014a}. \citet{Kopp2014a} synthesized these different sources to generate self-consistent, probabilistic projections of local sea-level changes around the world  under different future emission trajectories.

Combined with historical records of storm tides, RSL projections provide insight into the changes in expected flood frequencies over the 21st century. We summarize the RSL projections of \citet{Kopp2014a} for North Carolina and apply the method of \citet{Tebaldi2012a} and \citet{Kopp2014a} to calculate their implications for flood-return periods.

Note that the projections of \citet{Kopp2014a} are not identical to those of the expert assessment of the Intergovernmental Panel on Climate Change (IPCC)'s Fifth Assessment Report \citep{IPCCAR5WG1ch13}. The most significant difference arises from the use of a self-consistent framework for estimating a complete probability distribution of RSL change, not just the likely (67\% probability) GMSL projections of the IPCC. \citet{Kopp2014a} and the IPCC estimate similar but not identical likely 21st century GMSL rise (under RCP 8.5, 62--100 cm vs. 53--97 cm, respectively; under RCP 2.6, 37--65 cm vs. 28--60 cm).

\section{Holocene sea-level change in North Carolina}

\begin{figure}[btp]
   \centering
   \includegraphics[width=12cm]{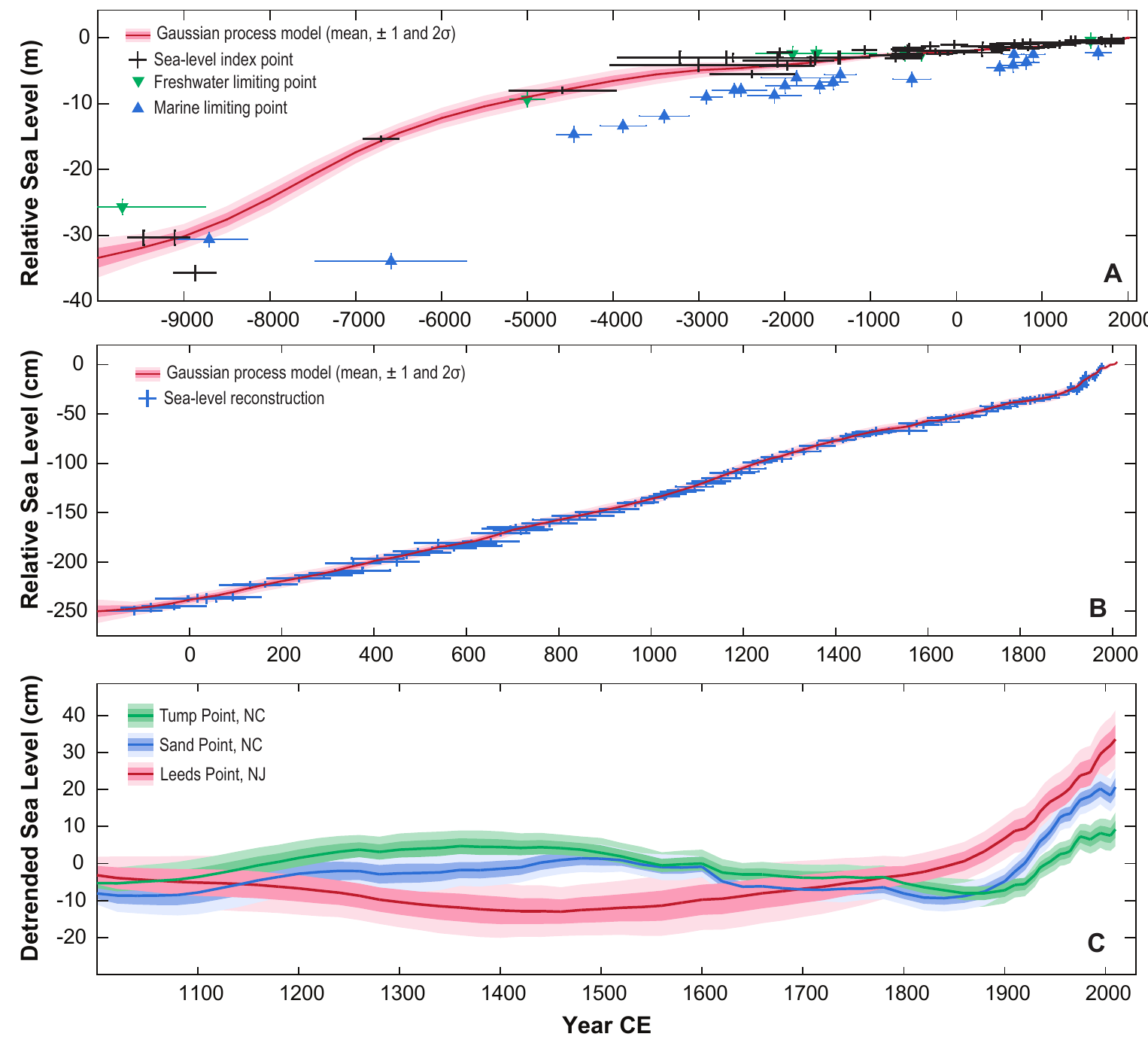}
   \caption{(a) Holocene RSL in North Carolina, showing a representative GP estimate for central North Carolina (\emph{red}), as well all index points (\emph{crosses}), marine limiting points (\emph{blue upward triangles}), and freshwater limiting points (\emph{green downward triangles}) from North Carolina. Index/limiting points shown with 2$\sigma$ error bars.   (b) RSL over the Common Era at Sand Point, North Carolina. (c) RSL detrended with respect to the 1000-1800 CE average rate for North Carolina (NC) and New Jersey (NJ).  GP estimates are shown with 1$\sigma$ (\emph{dark shading}) and 2$\sigma$ (\emph{light shading}) errors.
} \label{fig:timeseries} 
\end{figure}

\begin{figure}[btp]
   \centering
 \includegraphics[width=12cm]{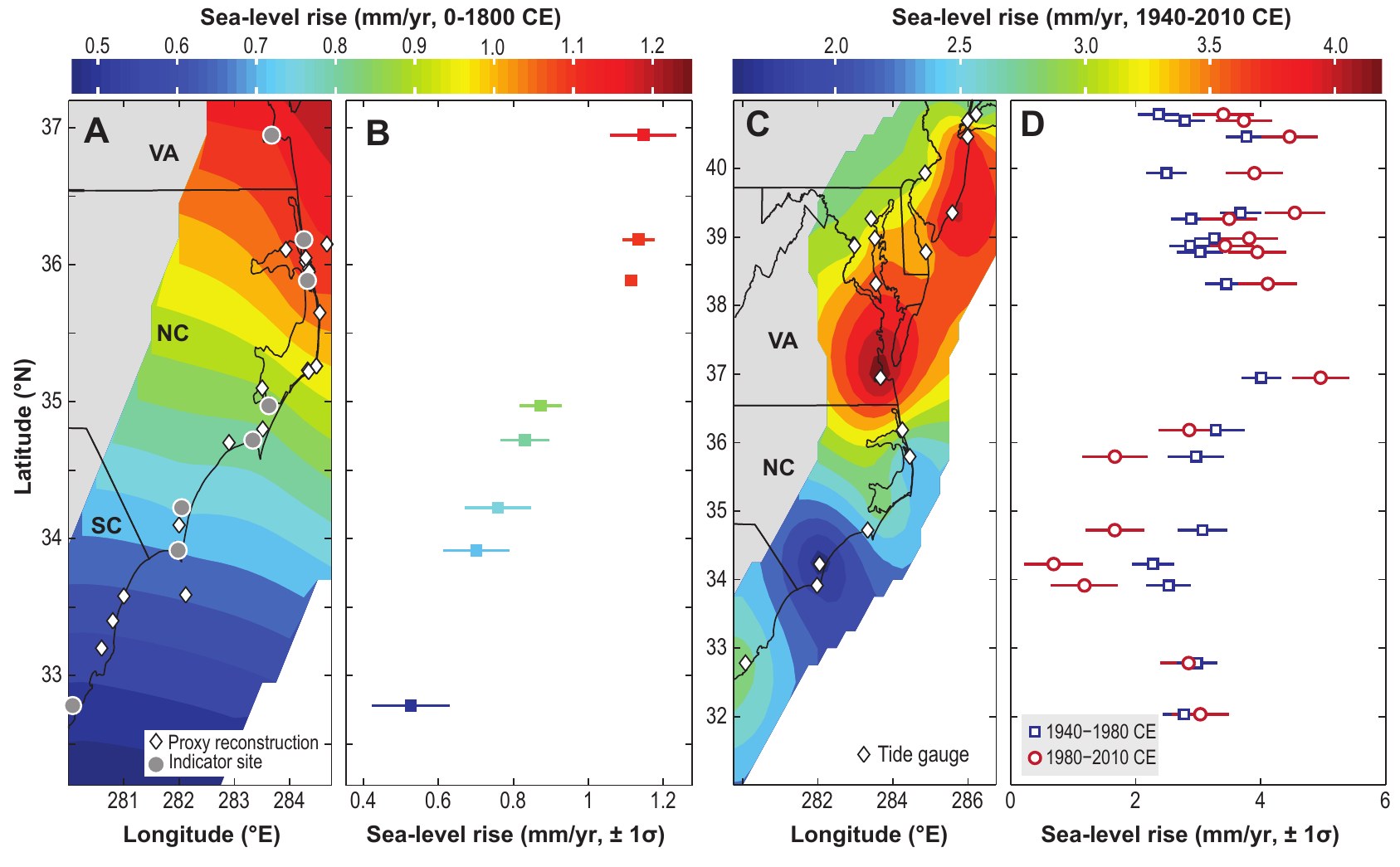}
   \caption{(a) Pre-Industrial Common Era rate of RSL rise  (0-1800 CE; mm/yr). Diamonds:  proxy sites; grey circles: selected tide gauges and continuous proxy records (as in Tables S-1 and S-2). Uncolored areas have $1\sigma$ uncertainty $\mathord{>}0.15$ mm/yr. (b) shows estimates at indicated tide-gauge and continuous proxy record sites (1$\sigma$ errors). (c) 1940-2010 rate of RSL rise. Diamonds: tide-gauge locations with $\mathord{>}60$ years of data. Uncolored areas have $1\sigma$ uncertainty $\mathord{>}0.5$ mm/yr. (d) 1940-1980 (blue squares) and 1980-2010 (red circles) rates of RSL rise at tide-gauge sites. 
 } \label{fig:ratemap} 
\end{figure}

RSL rose rapidly during the early and mid-Holocene, increasing in central North Carolina from -30.1 $\pm$ 1.8 m at 9000 BCE to -4.1 $\pm$ 0.7 m at 2000 BCE (Fig. \ref{fig:timeseries}a). The rate of RSL rise decreased over time, as a result of  declining input from shrinking land ice reservoirs and slowing GIA \citep{Peltier2004a,Milne2008a}, from a millennially-averaged rate of 6.8 $\pm$ 1.2 mm/yr at 8000 BCE to 0.8 $\pm$ 1.0 mm/yr at 2500 BCE. A declining GIA rate with increasing distance from the center of the Laurentide ice sheet \citep{Engelhart2009a}, along with a contribution from tectonic uplift along the Cape Fear Arch \citep{vandePlassche2014a}, caused spatial variability in the rate of Common Era RSL rise along the U.S. Atlantic coast and within North Carolina (Fig. \ref{fig:ratemap}a). At Sand Point in northern North Carolina, RSL rose from -2.38 $\pm$ 0.06 m at 0 CE to -0.37 $\pm$ 0.05 m by 1800 CE, an average rate of 1.11 $\pm$ 0.03 mm/yr. In the Wilmington area, the estimated average rate of RSL rise from 0 to 1800 CE was 0.8 $\pm$ 0.2 mm/yr (Fig. \ref{fig:ratemap}a-b; Table S-1).

Century-average rates of RSL change varied around these long-term means. For example, between 1000 and 1800 CE at Sand Point, century-average rates of RSL change ranged from a high of 1.7 $\pm$ 0.5 mm/yr (in the 12th century) to a low of 0.9 $\pm$ 0.5 mm/yr (in the 16th century) (Figure \ref{fig:timeseries}b).  Synchronous sea-level changes occurred in southern NC over the same period of time \citep{Kemp2011a}. However, the sign of the North Carolina RSL rate changes contrasts with that reconstructed at sites further north in New Jersey \citep{Kopp2013a} (Figure \ref{fig:timeseries}c). This contrast suggests a role for changes in ocean and atmosphere circulation, such as a shift in the position or strength of the Gulf Stream, in explaining these variations. A strengthening of the Gulf Stream (the opposite of the pattern depicted in Figure \ref{fig:map}d) would be consistent with the observations. The absence of similarly timed variations in Florida \citep{Kemp2014a} excludes a significant contribution from the static-equilibrium fingerprint of GrIS mass changes (Figure \ref{fig:map}c).

\section{Twentieth-century sea-level changes in North Carolina}

The most prominent feature in the North Carolina Common Era sea-level record is the acceleration of the rate of rise between the 19th and 20th centuries (Figure \ref{fig:timeseries}b-c). At Sand Point, the average rate of RSL rise over the 19th century (1.0 $\pm$ 0.5 mm/yr) was within the range of previous Common Era variability and close to the long-term average. By contrast, it is extremely likely ($P = 0.95$) that the 2.7 $\pm$ 0.5 mm/yr experienced in the 20th century was not exceeded in any century since at least the 10th century BCE (which had a rate of 1.2 $\pm$ 1.6 mm/yr). Average 20th century RSL rates  range from 2.1 $\pm$ 0.5 mm/yr at Wilmington  to 3.5 $\pm$ 0.3 mm/yr at Tump Point (Table S-1). 

Spatial patterns of sea-level variability are detectable at higher temporal frequencies in the  tide-gauge record \citep{Kopp2013a,Yin2013a} (Figure \ref{fig:ratemap}c-d; Table S-2). From 1940 to 1980 CE, sea-level rise in both North Carolina and the U.S. mid-Atlantic region exceeded the global mean.  At Wilmington and Duck, the average rates were 2.3 $\pm$ 0.7 mm/yr and $3.3 \pm 0.9$ mm/yr, respectively, compared to 2.8 $\pm$ 0.6 mm/yr at New York City and a GMSL rise of  $0.8 \pm 0.8$ mm/yr \citep{Hay2015a}. This pattern changed over the interval from 1980 to 2010 CE, when the rate of GMSL rise increased to $2.5 \pm 0.5$ mm/yr while rates of RSL rise south of Cape Hatteras remained stationary or decreased (1.7 $\pm$ 1.0 mm/yr at Beaufort, 0.7 $\pm$ 0.9 mm/yr at Wilmington, and 1.2 $\pm$ 1.1 mm/yr at Southport). In contrast, sites north of Cape Hatteras experienced a significant increase in rate; at New York City, for example, RSL rose at 3.7 $\pm$ 0.9 mm/yr. 

Several recent papers identified this regional phenomenon in the northeastern U.S. as a ``hot spot'' of sea-level acceleration \citep{Sallenger2012a,Boon2012a,Ezer2012a,Kopp2013a}. Less attention has been paid to its counterpart in the southeastern U.S., which might be regarded as a ``hot spot'' of deceleration, especially when considered in the context of the GMSL acceleration occurring over the same interval. The pattern of a sea-level increase north of Cape Hatteras and sea-level decrease south of Cape Hatteras is consistent with a northward migration of the Gulf Stream \citep{Yin2013a,Rahmstorf2015a}. It is also consistent with the dominant spatial pattern of change seen in the North Carolina and New Jersey proxy reconstructions from the 16th through the 19th century (Figure \ref{fig:timeseries}c). Dredging has, however, contaminated some  North Carolina tide gauges, rendering a simple assessment of the ocean dynamic contribution during the 20th century challenging.

\section{Future sea-level projections for North Carolina}

\begin{figure}[btp]
   \centering
  \includegraphics[width=12cm]{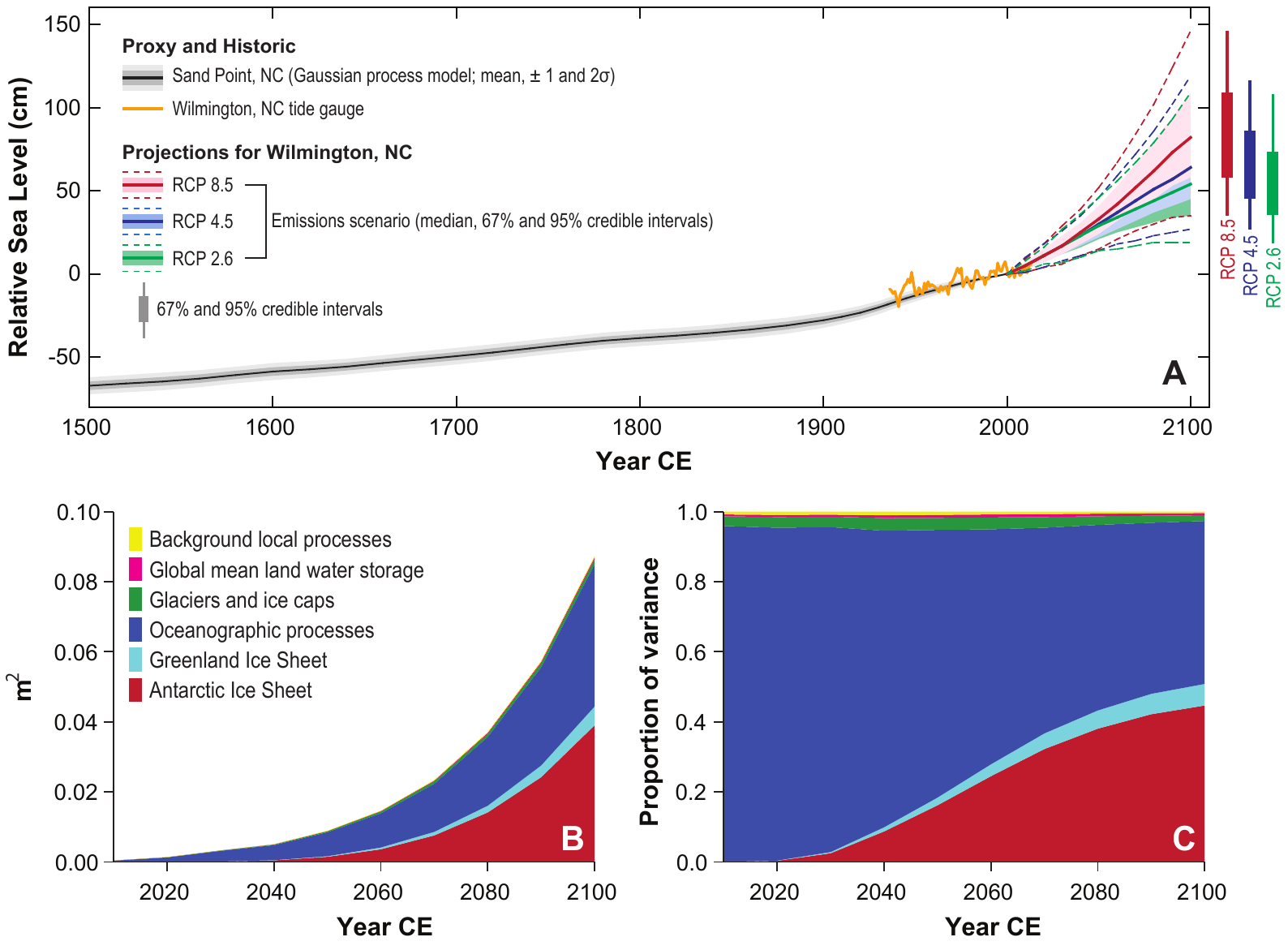}
   \caption{(a) GP estimate of sea-level at Sand Point (\emph{black}), annual Wilmington tide-gauge data (\emph{orange}),  and \citet{Kopp2014a} projections for RCP 8.5 (\emph{red}), 4.5 (\emph{blue}), and 2.6 (\emph{green}). Shading/dashed lines = 67\%/95\% credible intervals. Bars and whiskers represent 67\% and 95\% credible intervals of 2100 CE projections. All heights relative to 2000 CE. (b-c) Sources of uncertainty in RCP 8.5 20-year-average sea-level rise projection at Wilmington, shown in units of (b) variance and (c)  fractional variance as in \citet{Kopp2014a}. } \label{fig:proj} 
\end{figure}

\begin{table}[tbp]
{
\caption{Projected sea-level rise in North Carolina under RCP 8.5 and RCP 2.6}
\label{tab:proj}
\begin{tabular}{l | l l l l l |  l l l l  }
cm & \multicolumn{5}{c|}{RCP 8.5} & \multicolumn{4}{c}{RCP 2.6} \\
  & 50 &  17--83  &  5--95  &  0.5--99.5  & 99.9 & 50 &  17--83  &  5--95  &  0.5--99.5  \\ \hline
\multicolumn{10}{l}{DUCK, NC} \\
2030 & 23 & 16--29 & 12--33 & 6--39 & 43 & 22 & 17--28 & 12--32 & 7--38 \\
2050 & 41 & 31--51 & 24--59 & 15--72 & 83 & 37 & 28--46 & 22--53 & 13--66 \\
2100 & 100 & 73--129 & 54--154 & 29--214 & 304 & 70 & 50--93 & 36--113 & 17--181 \\
2150 & 160 & 124--206 & 103--255 & 76--425 & 627 & 99 & 71--136 & 56--184 & 39--357 \\
 2200 & 225 & 166--304 & 134--394 & 99--715 & 1055 & 131 & 80--196 & 58--287 & 33--607 \\
\hline
\multicolumn{10}{l}{WILMINGTON, NC} \\
2030 & 17 & 12--23 & 8--27 & 3--33 & 36 & 17 & 12--21 & 9--25 & 4--30 \\
2050 & 33 & 24--42 & 18--48 & 10--61 & 75 & 29 & 21--36 & 16--42 & 9--55 \\
2100 & 82 & 58--109 & 42--132 & 20--194 & 281 & 54 & 36--74 & 24--94 & 8--162 \\
2150 & 135 & 101--180 & 81--230 & 57--395 & 596 & 77 & 48--113 & 34--161 & 16--334 \\
2200 & 194 & 136--273 & 105--364 & 74--678 & 1016 & 101 & 50--166 & 27--257 & 3--575 \\
\hline
\multicolumn{10}{l}{Values represent two-decade averages and are in cm above 1990--2010 (`2000') mean sea level. } \\
\multicolumn{10}{l}{Columns correspond to different projection probabilities. For example, the  ``5-95'' columns}\\
\multicolumn{10}{l}{correspond to the 5th to 95th percentile; in IPCC  terms, the  `very likely' range.} \\
\multicolumn{10}{l}{The RCP 8.5 99.9th percentile corresponds to the maximum level physically possible.}
\end{tabular}
}
\end{table}

The integrated assessment and climate modeling communities  developed Representative Concentration Pathways (RCPs) to describe future emissions of greenhouse gases consistent with varied socio-economic and policy scenarios \citep{VanVuuren2011a}.  These pathways provide boundary conditions for projecting future climate and sea-level changes.  RCP 8.5 is consistent with high-end business-as-usual emissions. RCP 4.5 is consistent with moderate reductions in greenhouse gas emissions, while RCP 2.6 requires strong emissions reductions. These three RCPs respectively yield likely ($P = 0.67$) global mean temperature increases in 2081-2100 CE of 3.2--5.4$^\circ$C, 1.7--3.2$^\circ$C, and 0.9--2.3$^\circ$C above 1850-1900 CE levels \citep{IPCCAR5WG1ch12}.

A bottom-up assessment of the factors contributing to sea-level change \citep{Kopp2014a} indicates that, regardless of the  pathway of future emissions, it is virtually certain ($P > 0.998$) that both Wilmington and Duck will experience a RSL rise over the 21st century and very likely ($P > 0.90$) that the rate of that rise will exceed the rate observed during the 20th century. Below, we summarize the bottom-up projections of \citet{Kopp2014a} for Wilmington and Duck, NC, which bracket the latitudinal extent and degree of spatial variability across the state (Tables \ref{tab:proj}, S-3, S-4, S-5). 

Under the high-emissions RCP 8.5 pathway, RSL at Wilmington will very likely ($P = 0.90$) rise by 8--27 cm (median of 17 cm) between 2000 and 2030 CE and by 18--48 cm (median of 33 cm) between 2000 and 2050 CE (Figure \ref{fig:proj}a). Projected RSL rise varies modestly across the state, with a very likely rise of 12--33 cm (median 23 cm) between 2000 and 2030 CE  and of 24--59 cm (median of 41 cm) between 2000 and 2050 CE at Duck. Because sea level responds slowly to climate forcing, projected RSL rise before 2050 CE can be reduced only weakly ($\about$3-6 cm) through greenhouse gas mitigation. 

It is important to consider these numbers in the context of the background variability in annual-mean and decadal-mean RSL. Relative to 20-year-mean RSL, annual-mean RSL as measured by the Wilmington tide gauge has a standard deviation of $\about 8$ cm, so the median projection for 2030 CE is only slightly above twice the standard deviation. It would therefore not be surprising to see an isolated year with RSL as high as that projected for 2030 CE even in the absence of a long-term trend. However, consecutive years of that height would be unexpected, as decadal-mean RSL has a standard deviation of $\about 1$ cm. Given the magnitude of decadal variability, however, differences in projections of $\mathord{<}\about 4$ cm should not be viewed as significant.

Reductions in greenhouse gases over the course of the 21st century can significantly affect sea-level rise after 2050 CE. Under the high-emissions RCP 8.5 pathway, RSL at Wilmington is very likely to rise by 42--132 cm (median of 82 cm) between 2000 and 2100 CE, while under the low-emissions RCP 2.6 pathway, it is very likely to rise by 24--94 cm (median of 54 cm). The maximum physically possible 21st century sea-level rise is significantly higher ($\about$280 cm), although the estimated probability of such an outcome is extremely low ($P \approx 0.001$) \citep{Kopp2014a}. Projected RSL rise varies modestly across the state, with a very likely rise of 54--154 cm (median of 100 cm) under RCP 8.5 and 36--113 cm (median of 70 cm) under RCP 2.6 at Duck, a difference from Wilmington of $\about$12--22 cm.

Uncertainty in projected RSL rise in North Carolina stems from two main sources: the (1)  oceanographic and (2) Antarctic ice sheet responses to climate change. The former source dominates the uncertainty through most of the century, with the Antarctic response coming to play a roughly equal role by the end of the century (Figure \ref{fig:proj}b-c). At Wilmington, under RCP 8.5, ocean dynamics is likely ($P = 0.67$) to contribute -9 to +17 cm (median 5 cm) to 21st century sea-level rise. The dynamic contribution increases to the north, with -9 to +25 cm (median 8 cm) likely at Duck. These contributions are less than those in the northeastern United States; for example, at New York, ocean dynamics are likely to contribute -6 to +35 cm (median 14 cm).

The GrIS contribution to uncertainty in North Carolina RSL change is smaller than the Antarctic contribution because of two factors. First, GrIS makes a smaller overall contribution to GMSL uncertainty, because GrIS mass change is dominated by surface mass balance, while the behavior of WAIS is dominated by more complex and uncertain ocean/ice sheet dynamics.  Second, the GrIS contribution to North Carolina RSL change and to its uncertainty  is diminished by the static-equilibrium fingerprint effect to about 60\% of its global mean value.

\section{Implications of sea-level rise for flood risk and economic damages}

Based on historical storm tides, the `1-in-10 year' flood (i.e., the flood level with a probability of 10\% in any given year) at the Wilmington tide gauge is 0.60 m above current mean higher high water (MHHW). In the absence of sea-level rise, one would expect three such floods over a 30-year period. Assuming no increase in the height of storm-driven flooding relative to mean sea level and accounting for the probability distribution of projected sea-level rise as in \citet{Kopp2014a},  seven similar magnitude floods are expected between 2000 and 2030 (regardless of RCP). Between 2000 and 2050, the expected number of years experiencing a flood at 0.60 m  above current MHHW increases from 5 to 21. After 2050, regardless of RCP, almost every year is expected to see at least one flood at 0.60 m  above current MHHW. Similarly, the expected number of 0.93 m `1-in-100 year' floods will increase with projected sea-level rise. The `1-in-100 year' flood  is expected about 1.6--1.8 times between 2000 and 2050 (rather than the 0.5 times expected in the absence of sea-level rise). During the second half of the century, `1-in-100 year' flooding is expected in 29 of 50 years under RCP 8.5 and 17 of 50 years under RCP 2.6.

\citet{Houser2015a} characterized the costs of projected sea-level rise and changes in flood frequency using the Risk Management Solutions North Atlantic Hurricane Model, which models wind and coastal flood damage to property and interrupted businesses caused by a database of tens of thousands of synthetic storm events. Under all RCPs,  projected RSL rise in North Carolina would likely ($P = 0.67$) place $\mathord{>}\$4$ billion of current property below MHHW by 2050 and $\mathord{>}\$17$ billion by 2100. Statewide (assuming fixed distribution and value of property), average annual insurable losses from coastal storms will very likely ($P = 0.90$) increase by 4-17\% between 2011 and 2030 and by 16-75\% between 2011 and 2050 (regardless of RCP). By 2100, they are very likely to increase by 50-160\% under RCP 8.5 and 20-150\% under RCP 2.6 \citep{Houser2015a}. Projected increases in the intensity of tropical cyclones under RCP 8.5 \citep{Emanuel2013a} may amplify the increase in losses by  $\about$1.5x by 2050 and $\about$2.1x by 2100. These cost estimates assume a fixed distribution and valuation of property; intensification of development along the coastline will increase exposure and therefore cost, while protective measures will decrease exposure and cost.

\section{Concluding remarks}

North Carolina Session Law 2012-202/House Bill 819 requires assessment of   future sea-level change trajectories that include ``sea-level fall, no movement in sea level, deceleration of sea-level rise, and acceleration of sea-level rise.'' Geological and historical records indicate that, over the last 11,000 years, North Carolina  experienced periods of RSL deceleration and acceleration, but no periods of RSL stasis or fall. 

\begin{itemize}

\item 	Millennially-averaged  RSL rise in central North Carolina decelerated from  8000 BCE (6.8 $\pm$ 1.2 mm/yr) until 2500 BCE (0.8 $\pm$ 1.0 mm/yr). 
\ \\

\item From 0 to 1800 CE, average RSL rise rates within North Carolina varied  from 1.11 $\pm$ 0.03 mm/yr in northern North Carolina to 0.8 $\pm$ 0.2 mm/yr in southern North Carolina (in the vicinity of the Cape Fear Arch, and farther away from the peripheral bulge). Century-average rates of sea-level change varied around these long-term means. Comparison of records  along the U.S. Atlantic coast  indicate that pre-Industrial Common Era sea-level accelerations and decelerations had a spatial pattern consistent with variability in the strength and/or position of the Gulf Stream.
\ \\

\item It is extremely likely ($P = 0.95$) that the accelerated rate of 20th century RSL rise at Sand Point, NC, (2.7 $\pm$ 0.5 mm/yr)  had not been reached in any century since at least the 10th century BCE.
\ \\

\item Between 1940-1980 and 1980-2010, sea level in North Carolina decelerated relative to the global mean and possibly in absolute terms (at Wilmington, from 2.3 $\pm$ 0.5 mm/yr to 0.7 $\pm$ 0.9 mm/yr; at Southport, from 2.5 $\pm$ 0.7 mm/yr to 1.2 $\pm$ 1.1 mm/yr), while sea-level rise accelerated north of Cape Hatteras. The spatial pattern and the magnitude of change are consistent with Gulf Stream variability.
\ \\

\item It is virtually certain ($P = 0.99$) that RSL rise at Wilmington between 2000 and 2050 will exceed 2.2 mm/yr, nearly three times the 0-1800 CE average rate. It is extremely likely ($P = 0.95$) that it will exceed 3.2 mm/yr, in excess of the 20th century average of 2.2 $\pm$ 0.6 mm/yr. Under the high-emissions RCP 8.5 pathway, RSL is very likely to rise by 42--132 cm, and under the low-emissions RCP 2.6 pathway RSL is very likely to rise by 24--94 cm between 2000 and 2100.\ \\

\item Storm flooding in North Carolina  will be increasingly exacerbated by sea-level rise. After 2050, the current `1-in-10 year' flood is expected to occur in Wilmington almost every year and the `1-in-100 year' flood is expected to occur in about 17--29 years. Assuming the current distribution of property and economic activity, average annual insurable losses statewide would very likely increase by 50-160\% under RCP 8.5 and 20-150\% under RCP 2.6. 

\end{itemize}

\begin{acknowledgements}
We thank the American Climate Prospectus research team for assisting with the development of the sea-level rise projections, E. Morrow for retrieving the CESM ocean dynamic sea-level change from the CMIP5 archive, C. Zervas for assistance with the NC tide-gauge data, and C. Hay for helpful comments. Funding was provided by the Risky Business Project, National Science Foundation awards EAR-1052848, ARC-1203415, EAR-1402017, and OCE-1458904, National Oceanic \& Atmospheric Administration grant
NA11OAR4310101, and New Jersey Sea Grant project 6410-0012. C. Tebaldi is supported by the Regional and Global Climate
Modeling Program of the U.S. Department of Energy's, Office of Science (BER),  Cooperative Agreement DE-FC02-97ER62402. This paper is a contribution to International Geoscience Program project 588
`Preparing for coastal change' and the PALSEA2 (Palaeo-Constraints on Sea-Level Rise) project of Past Global Changes/IMAGES (International Marine Past Global Change Study).
\end{acknowledgements}


\newpage
\clearpage

\iftoggle{supplement}{

\clearpage
\newpage

\setcounter{page}{1}

\renewcommand{\thepage}{S-\arabic{page}}

\setcounter{figure}{0}
\renewcommand{\thefigure}{S-\arabic{figure}}

\setcounter{equation}{0}
\renewcommand{\theequation}{S-\arabic{equation}}

\setcounter{table}{0}
\renewcommand{\thetable}{S-\arabic{table}}

\renewcommand{\thesection}{S}

\section*{Supporting Information:  Spatio-temporal statistical model}

The spatio-temporal sea-level field $f(\mathbf{x},t)$ is modeled as a sum of Gaussian processes \citep{Rasmussen2006a} with different characteristic spatial and temporal scales.

\begin{align}
f(\mathbf{x},t) & = l(\mathbf{x},t) + m(\mathbf{x},t) + h(\mathbf{x},t) 
\end{align}

\noindent Each field has a prior mean of zero and spatially and temporally separable prior covariances given by

\begin{align}
k_l(\mathbf{x}_1,t_1,\mathbf{x}_2,t_2) & = \sigma_l^2  \cdot C_\frac{3}{2}(|t_2-t_1|, \tau_l ) \cdot C_\frac{5}{2}(r(\mathbf{x}_1,\mathbf{x}_2), \gamma_l) \\
k_m(\mathbf{x}_1,t_1,\mathbf{x}_2,t_2) & = \sigma_m^2 \cdot C_\frac{3}{2}(|t_2-t_1|, \tau_m ) \cdot C_\frac{1}{2}(r(\mathbf{x}_1,\mathbf{x}_2), \gamma_m) \\
k_h(\mathbf{x}_1,t_1,\mathbf{x}_2,t_2) & = \sigma_h^2  \cdot C_\frac{3}{2}(|t_2-t_1|, \tau_h) \cdot C_\frac{1}{2}(r(\mathbf{x}_1,\mathbf{x}_2), \gamma_m) \\
\end{align}

\noindent where $C_\nu(r,\lambda)$ is a Mat\'ern covariance function with scale $\lambda$ and smoothness parameter $\nu$. Here $\sigma_i^2$ are the amplitudes of the prior variances, $\tau_i$ are characteristic time scales, $\gamma_i$ are characteristic length scales, and $r(\mathbf{x}_1,\mathbf{x}_2)$ is the angular distance between $\mathbf{x}_1$ and $\mathbf{x}_2$.

The observations $y(\mathbf{x},t')$ are modeled as

\begin{align}
y(\mathbf{x},t') & =  f(\mathbf{x},t+\epsilon_t) + w(\mathbf{x},t') + \epsilon_y + y_0(\mathbf{x}),
\end{align}

\noindent where $t'$ is the true age of the observation, $t$ the mean observed age, $w$ a process that captures sea-level variability at a sub-decadal level (which we treat here as noise), $\epsilon_t$ and $\epsilon_y$ are errors in the age and sea-level observations, and $y_0$ is a site-specific datum offset. For tide gauges, $\epsilon_t$ is zero and $\epsilon_y$ is estimated during a smoothing process (see below) in which annual data are assumed to have uncorrelated, normally distributed noise with standard deviation 3 mm. For proxy data, $\epsilon_t$ and $\epsilon_y$ are treated as independent and normally distributed, with a standard deviation specified for each observation based on the original publication. The sub-decadal  and datum offset processes are modeled as Gaussian processes with mean zero and prior covariances given by

\begin{align}
k_w(\mathbf{x}_1,t_1,\mathbf{x}_2,t_2) & = \sigma_w^2 \delta(t_1,t_2) \delta(\mathbf{x}_1,\mathbf{x}_2) \\
k_{0}(\mathbf{x}_1,\mathbf{x}_2) & = \sigma_{0}^2 \delta(\mathbf{x}_1,\mathbf{x}_2),
\end{align}

\noindent where $\delta(\mathbf{x}_1,\mathbf{x}_2)$ is the Kronecker delta function.
Geochronological uncertainties are incorporated using the noisy-input Gaussian process method of \citet{McHutchon2011a}:

\begin{align}
y(\mathbf{x},t') & \approx f(\mathbf{x},t') + \epsilon_t  f'(\mathbf{x},t')+  w(\mathbf{x},t) +  \epsilon_y + y_0(\mathbf{x}).
\end{align}

The low-frequency process $l(\mathbf{x},t)$ (physically corresponding to GIA, tectonics, long-term sediment compaction, and long-term GMSL change), medium-frequency process $m(\mathbf{x},t)$, and high-frequency process $h(\mathbf{x},t)$ all have Mat\'ern temporal covariance functions with smoothness parameter $\nu = 1.5$, implying a functional form in which the first derivative is everywhere defined. The low-frequency process is assumed to vary smoothly over space ($\nu = 2.5$), while the medium- and high-frequency process are allowed to vary more roughly ($\nu = 0.5$). The length scale $\gamma_m$ is required to be equal for the medium- and high-frequency processes, as both are expected to reflect similar oceanographic processes operating on different timescales.

The hyperparameters  $\mathbf{\Theta} = \{\sigma_l, \sigma_m, \sigma_h, \sigma_w, \sigma_0, \tau_l, \tau_m, \tau_w, \gamma_l, \gamma_m\}$   are set through a three-step optimization process. First, the hyperparameters of a simplified model, in which a linear term replaces the low-frequency process, are globally optimized through simulated annealing to maximize the marginal likelihood $\mathcal{L}(\mathbf{\Theta}|\mathbf{y}_1)$, where $\mathbf{y}_1$ is the set of post-1000 BCE observations. Second,  the hyperparameters of $m(\mathbf{x},t)$, $h(\mathbf{x},t)$ and $w(\mathbf{x},t)$ are fixed. The remaining hyperparameters of the full model -- the amplitude, scales, and spatial roughness of the low-frequency process, as well as the datum offset -- are globally optimized so as to maximize the marginal likelihood $\mathcal{L}(\mathbf{\Theta}|\mathbf{y}_2)$, where $\mathbf{y}_2$ is the complete data set . Finally, all the hyperparameters are locally optimized to maximize the marginal likelihood  $\mathcal{L}(\mathbf{\Theta}|\mathbf{y}_2)$. This multi-step process improves performance relative to globally optimizing all hyperparameters simultaneously and is guided by the recognition that the long-term, low-resolution data provide the greatest insight into the lowest-frequency processes while the salt-marsh and tide-gauge data provide the greatest insight into the medium-frequency and high-frequency processes. The optimized time scales of the high-, medium- and low-frequency processes are respectively  $\tau_l = 14.5$ kyr, $\tau_m = 296$ years and $\tau_h =  6.3$ years; other hyperparameters are shown in Table \ref{Stab:hyperparameters}. 

Annual mean tide-gauge data are decadally averaged prior to incorporation into the analysis. To accommodate data gaps estimate the covariance of the decadal averages, we fit each annual record $y_j(t)$ separately with the model
\begin{align}
y_j(t) = \alpha_j (t-t_0) + d_j(t) + y_{0,j},
\end{align}
where $\alpha_j$ is a slope, $t_0$ a reference time period, and $d_j(t)$  a Gaussian process with prior mean zero and a prior Mat\'ern covariance. Hyperparameters are optimized on a site-by-site basis to maximize their marginal likelihood. Decadal averages, including their covariances, are then taken from the interpolated process $y_j(t)$.

\begin{table}[p]
{\scriptsize
\caption{Common Era sea-level rates (mm/yr)}
\label{Stab:linratesCE}

\begin{tabular}{l | l l |  l |  l l l l }
Site & Lat & Long & 0-1800 & 1000-1500 & 1500-1800 & 1800-1900 & 1900-2000 \\ \hline
GMSL & & & & & & & $1.3 \pm 0.2$ \\ \hline
New York, NY & 40.7 & -74.0 & 1.69 $\pm$ 0.18 & 1.5 $\pm$ 0.5 & 1.9 $\pm$ 0.7 & 2.1 $\pm$ 0.7 & 2.9 $\pm$ 0.3 \\
Leeds Point, NJ & 39.5 & -74.4 & 1.52 $\pm$ 0.09 & 1.2 $\pm$ 0.2 & 1.7 $\pm$ 0.4 & 2.4 $\pm$ 0.8 & 3.8 $\pm$ 0.5 \\
Cape May, NJ & 39.1 & -74.8 & 1.46 $\pm$ 0.10 & 1.2 $\pm$ 0.2 & 1.5 $\pm$ 0.3 & 2.2 $\pm$ 0.6 & 3.7 $\pm$ 0.5 \\
Sewell's Point, VA & 37.0 & -76.3 & 1.15 $\pm$ 0.18 & 1.2 $\pm$ 0.5 & 0.9 $\pm$ 0.6 & 1.6 $\pm$ 0.9 & 4.2 $\pm$ 0.5 \\
Duck, NC & 36.2 & -75.8 & 1.13 $\pm$ 0.08 & 1.4 $\pm$ 0.3 & 1.0 $\pm$ 0.4 & 1.2 $\pm$ 0.6 & 3.1 $\pm$ 0.6 \\
Sand Point, NC & 35.9 & -75.7 & 1.11 $\pm$ 0.03 & 1.4 $\pm$ 0.1 & 1.0 $\pm$ 0.2 & 1.0 $\pm$ 0.5 & 2.7 $\pm$ 0.5 \\
Oregon Inlet, NC & 35.8 & -75.6 & 1.11 $\pm$ 0.07 & 1.4 $\pm$ 0.2 & 1.0 $\pm$ 0.3 & 1.1 $\pm$ 0.6 & 2.6 $\pm$ 0.5 \\
Tump Point, NC & 35.0 & -76.4 & 0.87 $\pm$ 0.11 & 1.2 $\pm$ 0.2 & 0.7 $\pm$ 0.2 & 1.4 $\pm$ 0.4 & 3.5 $\pm$ 0.3 \\
Beaufort, NC & 34.7 & -76.7 & 0.83 $\pm$ 0.13 & 1.2 $\pm$ 0.3 & 0.7 $\pm$ 0.4 & 1.2 $\pm$ 0.7 & 2.9 $\pm$ 0.5 \\
Wilmington, NC & 34.2 & -78.0 & 0.76 $\pm$ 0.18 & 1.0 $\pm$ 0.5 & 0.7 $\pm$ 0.6 & 0.9 $\pm$ 1.0 & 2.1 $\pm$ 0.5 \\
Southport, NC & 33.9 & -78.0 & 0.70 $\pm$ 0.18 & 0.9 $\pm$ 0.5 & 0.6 $\pm$ 0.6 & 0.9 $\pm$ 1.0 & 2.3 $\pm$ 0.6 \\
Charleston, SC & 32.8 & -79.9 & 0.53 $\pm$ 0.21 & 0.6 $\pm$ 0.6 & 0.4 $\pm$ 0.7 & 1.1 $\pm$ 1.1 & 2.9 $\pm$ 0.5 \\
Fort Pulaski, GA & 32.0 & -80.9 & 0.47 $\pm$ 0.19 & 0.5 $\pm$ 0.5 & 0.3 $\pm$ 0.7 & 1.0 $\pm$ 1.1 & 2.7 $\pm$ 0.5 \\
Nassau, FL & 30.6 & -81.7 & 0.41 $\pm$ 0.05 & 0.5 $\pm$ 0.2 & 0.4 $\pm$ 0.3 & 0.7 $\pm$ 0.8 & 1.9 $\pm$ 0.4 \\

\hline
\multicolumn{7}{l}{\tiny{Errors are $\pm2\sigma$. GMSL from  \citet{Hay2015a}.}}

\end{tabular}
}
\end{table}

\begin{table}[p]
{\scriptsize
\caption{Industrial era sea-level rates (mm/yr)}
\label{Stab:linrates20th}

\begin{tabular}{l | l l |  l l l l }
Site & Lat & Long & 1860-1900 &1900-1940 &1940-1980 &1980-2010 \\ \hline
GMSL & & & & $1.2 \pm 1.1$ & $0.8 \pm 0.8$ & $2.5 \pm 0.5$ \\ \hline
New York, NY & 40.7 & -74.0 & 2.5 $\pm$ 0.7 & 2.7 $\pm$ 0.7 & 2.8 $\pm$ 0.6 & 3.7 $\pm$ 0.9 \\
Atlantic City, NJ & 39.4 & -74.4 & 3.0 $\pm$ 1.1 & 3.7 $\pm$ 0.9 & 3.7 $\pm$ 0.7 & 4.6 $\pm$ 1.0 \\
Cape May, NJ & 39.1 & -74.8 & 2.8 $\pm$ 1.0 & 3.4 $\pm$ 0.9 & 3.4 $\pm$ 0.8 & 4.4 $\pm$ 1.1 \\
Sewell's Point, VA & 37.0 & -76.3 & 2.3 $\pm$ 1.3 & 3.9 $\pm$ 1.1 & 4.0 $\pm$ 0.6 & 5.0 $\pm$ 0.9 \\
Duck, NC & 36.2 & -75.8 & 1.7 $\pm$ 1.1 & 3.2 $\pm$ 1.0 & 3.3 $\pm$ 0.9 & 2.9 $\pm$ 1.0 \\
Sand Point, NC & 35.9 & -75.7 & 1.4 $\pm$ 1.0 & 3.0 $\pm$ 0.9 & 3.0 $\pm$ 0.8 & 2.0 $\pm$ 1.1 \\
Oregon Inlet, NC & 35.8 & -75.6 & 1.5 $\pm$ 1.0 & 3.0 $\pm$ 0.9 & 3.0 $\pm$ 0.9 & 1.7 $\pm$ 1.1 \\
Tump Point, NC & 35.0 & -76.4 & 2.0 $\pm$ 0.9 & 4.0 $\pm$ 0.8 & 3.7 $\pm$ 0.7 & 2.0 $\pm$ 1.1 \\
Beaufort, NC & 34.7 & -76.7 & 1.7 $\pm$ 1.1 & 3.5 $\pm$ 1.0 & 3.1 $\pm$ 0.8 & 1.7 $\pm$ 1.0 \\
Wilmington, NC & 34.2 & -78.0 & 1.3 $\pm$ 1.3 & 2.5 $\pm$ 1.2 & 2.3 $\pm$ 0.7 & 0.7 $\pm$ 0.9 \\
Southport, NC & 33.9 & -78.0 & 1.4 $\pm$ 1.4 & 2.5 $\pm$ 1.2 & 2.5 $\pm$ 0.7 & 1.2 $\pm$ 1.1 \\
Charleston, SC & 32.8 & -79.9 & 1.7 $\pm$ 1.5 & 2.8 $\pm$ 1.1 & 3.0 $\pm$ 0.7 & 2.9 $\pm$ 0.9 \\
Fort Pulaski, GA & 32.0 & -80.9 & 1.5 $\pm$ 1.4 & 2.4 $\pm$ 1.2 & 2.8 $\pm$ 0.7 & 3.0 $\pm$ 0.9 \\
Fernandina Beach, FL & 30.7 & -81.5 & 1.2 $\pm$ 1.3 & 1.5 $\pm$ 0.7 & 1.9 $\pm$ 0.7 & 2.3 $\pm$ 0.9 \\
\hline
\multicolumn{7}{l}{\tiny{Errors are $\pm2\sigma$. GMSL from  \citet{Hay2015a}.}}
\end{tabular}
}
\end{table}

\begin{table}[p]
{\scriptsize
\caption{Projected sea-level rise in North Carolina by decade under RCPs 8.5 and 2.6}
\label{tab:projext}
\begin{tabular}{l | l l l l l |  l l l l  }
cm & \multicolumn{5}{c}{RCP 8.5} & \multicolumn{4}{|c}{RCP 2.6} \\
  & 50 &  17--83  &  5--95  &  0.5--99.5  & 99.9 & 50 &  17--83  &  5--95  &  0.5--99.5  \\ \hline
\multicolumn{10}{l}{DUCK, NC} \\
2010 & 7 & 5--9 & 4--10 & 1--12 & 13 & 7 & 5--9 & 3--11 & 1--13 \\
2020 & 14 & 11--18 & 8--21 & 4--25 & 27 & 15 & 11--18 & 9--21 & 5--24 \\
2030 & 23 & 16--29 & 12--33 & 6--39 & 43 & 22 & 17--28 & 12--32 & 7--38 \\
2040 & 31 & 24--39 & 18--45 & 11--53 & 60 & 30 & 22--37 & 17--43 & 10--51 \\
2050 & 41 & 31--51 & 24--59 & 15--72 & 83 & 37 & 28--46 & 22--53 & 13--66 \\
2060 & 52 & 40--65 & 32--74 & 20--93 & 120 & 44 & 33--57 & 25--66 & 13--85 \\
2070 & 64 & 49--80 & 39--92 & 24--118 & 158 & 51 & 38--65 & 28--77 & 15--103 \\
2080 & 76 & 57--95 & 45--111 & 27--146 & 201 & 57 & 43--74 & 32--87 & 17--125 \\
2090 & 88 & 66--112 & 51--132 & 30--179 & 250 & 63 & 46--83 & 34--100 & 18--151 \\
2100 & 100 & 73--129 & 54--154 & 29--214 & 304 & 70 & 50--93 & 36--113 & 17--181 \\
\hline
2150 & 160 & 124--206 & 103--255 & 76--425 & 627 & 99 & 71--136 & 56--184 & 39--357 \\
2200 & 225 & 166--304 & 134--394 & 99--715 & 1055 & 131 & 80--196 & 58--287 & 33--607 \\
\hline \hline
\multicolumn{10}{l}{WILMINGTON, NC} \\
2010 & 5 & 3--7 & 2--8 & 0--10 & 11 & 5 & 4--7 & 2--8 & 1--10 \\
2020 & 11 & 8--15 & 5--17 & 1--21 & 22 & 11 & 8--14 & 6--16 & 4--18 \\
2030 & 17 & 12--23 & 8--27 & 3--33 & 36 & 17 & 12--21 & 9--25 & 4--30 \\
2040 & 25 & 18--31 & 13--36 & 6--44 & 51 & 23 & 17--29 & 12--34 & 6--42 \\
2050 & 33 & 24--42 & 18--48 & 10--61 & 75 & 29 & 21--36 & 16--42 & 9--55 \\
2060 & 42 & 31--53 & 24--62 & 13--80 & 107 & 34 & 25--44 & 18--52 & 9--70 \\
2070 & 52 & 39--66 & 29--78 & 17--103 & 142 & 39 & 28--51 & 20--61 & 9--88 \\
2080 & 62 & 46--79 & 35--94 & 19--130 & 183 & 44 & 31--58 & 23--71 & 10--111 \\
2090 & 73 & 53--94 & 40--113 & 21--162 & 229 & 49 & 34--66 & 24--82 & 10--135 \\
2100 & 82 & 58--109 & 42--132 & 20--194 & 281 & 54 & 36--74 & 24--94 & 8--162 \\
\hline
2150 & 135 & 101--180 & 81--230 & 57--395 & 596 & 77 & 48--113 & 34--161 & 16--334 \\
2200 & 194 & 136--273 & 105--364 & 74--678 & 1016 & 101 & 50--166 & 27--257 & 3--575 \\
\hline
\hline
\multicolumn{10}{l}{Values represent two-decade averages and are in cm above 1990--2010 (`2000') mean sea level. } \\
\multicolumn{10}{l}{Columns correspond to different projection probabilities. For example, the  ``5-95'' columns}\\
\multicolumn{10}{l}{correspond to the 5th to 95th percentile; in IPCC  terms, the  `very likely' range.} \\
\multicolumn{10}{l}{The RCP 8.5 99.9th percentile corresponds to the maximum level physically possible.}
\end{tabular}
}
\end{table}

\begin{table}[p]
{\scriptsize
\caption{Projected sea-level rise in North Carolina by decade under RCP 4.5}
\label{Stab:proj45}
\begin{tabular}{l | l l l l  }
cm &  \multicolumn{4}{|c}{RCP 4.5} \\
  & 50 &  17--83  &  5--95  &  0.5--99.5  \\ \hline
\multicolumn{5}{l}{DUCK, NC} \\
2010 & 7 & 5--9 & 3--11 & 1--13  \\
2020 & 14 & 11--18 & 8--21 & 4--25  \\
2030 & 22 & 17--27 & 13--31 & 8--36 \\
2040 & 30 & 24--37 & 19--42 & 13--50 \\
2050 & 39 & 30--47 & 23--54 & 15--67 \\
2060 & 47 & 36--59 & 28--68 & 17--86  \\
2070 & 56 & 42--71 & 32--82 & 18--108  \\
2080 & 64 & 48--82 & 37--96 & 21--130  \\
2090 & 72 & 54--93 & 41--110 & 23--158  \\
2100 & 81 & 60--105 & 45--126 & 25--188  \\
\hline
2150 & 121 & 84--164 & 60--209 & 30--374  \\
2200 & 160 & 101--232 & 67--315 & 24--618  \\
\hline
\hline
\multicolumn{5}{l}{WILMINGTON, NC} \\
2010 & 5 & 3--7 & 1--9 & -1--11 \\
2020 & 11 & 7--14 & 5--17 & 1--20 \\
2030 & 17 & 12--21 & 9--24 & 5--29 \\
2040 & 23 & 17--29 & 13--33 & 8--40 \\
2050 & 30 & 22--37 & 17--43 & 10--55 \\
2060 & 37 & 27--47 & 20--55 & 11--72 \\
2070 & 44 & 32--56 & 24--66 & 12--91 \\
2080 & 51 & 37--66 & 27--78 & 14--114 \\
2090 & 57 & 41--75 & 30--91 & 16--140 \\
2100 & 64 & 45--86 & 33--105 & 16--170 \\
\hline
2150 & 96 & 62--137 & 40--182 & 14--344 \\
2200 & 128 & 71--199 & 39--282 & 0--581 \\
\hline\hline
\multicolumn{5}{l}{\tiny Values in cm above 1990--2010 mean sea level. } \\
\multicolumn{5}{l}{\tiny Columns correspond to different probability ranges.} \\
\end{tabular}
}
\end{table}

\begin{table}[p]
{\scriptsize
\caption{Projected contributions to sea-level rise at Wilmington, NC, in 2100 CE}
\label{Stab:projcomp}
\begin{tabular}{l | l l l l l |  l l l l  }
cm & \multicolumn{5}{c}{RCP 8.5} & \multicolumn{4}{|c}{RCP 2.6} \\
 & 50 &  17--83  &  5--95  &  0.5--99.5  & 99.9 & 50 &  17--83  &  5--95  &  0.5--99.5   \\ \hline
Oc & 41 &  23--61  &  10--74  &  -10--93  &  100  & 21 &  8--34  &  -1--44  &  -15--57   \\
GrIS & 9 &  5--16  &  3--25  &  2--44  &  60  & 4 &  2--7  &  2--11  &  1--20   \\
AIS & 4 &  -8--18  &  -12--38  &  -15--109  &  180  & 7 &  -4--20  &  -8--40  &  -11--111  \\
GIC & 16 &  12--19  &  10--21  &  6--25  &  25  & 10 &  8--13  &  6--15  &  3--18  \\
LWS & 5 &  3--7  &  2--8  &  0--11  &  10  & 5 &  3--7  &  2--8  &  0--11   \\
Bkgd & 5 &  3--6  &  2--8  &  0--10  &  10  & 5 &  3--6  &  2--8  &  0--10  \\ \hline
Sum  & 82 &  58--109  &  42--132  &  20--194  &  280  & 54 &  36--74  &  24--94  &  8--162   \\
   \hline \multicolumn{10}{l}{\tiny{Oc: Oceanographic. GrIS: Greenland ice sheet. AIS: Antarctic ice sheet.   } }   \\
\multicolumn{10}{l}{\tiny{GIC: Glaciers and ice caps. LWS: Land water storage. Bkgd: Background.   } }   \\ 
\multicolumn{10}{l}{\tiny{All values are cm above 1990--2010 CE baseline. Columns correspond to probability ranges. } }   \\ 

\end{tabular}
}
\end{table}

\begin{table}[btp]
\caption{Optimized hyperparameters }
\label{Stab:hyperparameters}
\begin{tabular}{llll}
\hline
\multicolumn{2}{l}{\textbf{Low frequency}} \\
amplitude & $\sigma_l$ &  19.1 & m \\
time scale & $\tau_l$ & 14.5 & kyr \\
length scale & $\gamma_l$ & 25.0 & degrees \\
\hline
\multicolumn{2}{l}{\textbf{Medium frequency}} \\
amplitude & $\sigma_m$ & 119 & mm \\
time scale & $\tau_m$ & 296 & yr \\
length scale & $\gamma_m$ & 3.0 & degrees \\
\hline
\multicolumn{2}{l}{\textbf{High frequency}} \\
amplitude & $\sigma_h$ & 13.7 & mm \\
time scale & $\tau_h$ & 6.3 & y \\
length scale & $\gamma_m$ & 3.0 & degrees \\
\hline
\textbf{White noise} &  $\sigma_w$ & 4.2 & mm \\
\textbf{Datum offset} & $\sigma_0$ & 45 & mm \\
\hline
\end{tabular}

\end{table}

}

 \end{document}